\documentclass[article,12pt,showkeys,showpacs]{iopart}

\usepackage{graphicx}
\usepackage{amsfonts}

\begin{document}

\title[]{Deterministic SWAP gate using shortcuts to adiabatic passage}

\author{Yan Liang, Xin Ji, Hong-Fu Wang and Shou Zhang}

\address{Department of Physics, College of Science, Yanbian
University, Yanji, Jilin 133002, People's Republic of China}
\ead{jixin@ybu.edu.cn} \vspace{10pt}

\begin{abstract}
We theoretically propose an alternative
 method to realize a deterministic SWAP gate using
 shortcuts to adiabatic passage based on the approach
  of Lewis-Riesenfeld invariants in cavity
quantum electronic dynamics (QED). By combining Lewis-Riesenfeld
invariants with quantum Zeno dynamics,
 the SWAP gate can be achieved deterministically. The strict numerical results
show that our scheme is a fast and robust approach to achieve SWAP
gate.

\end{abstract}

%

\section{Introduction}

With the development of quantum science, the quantum computer
replaces the universal computer just around the corner. The
quantum logic gate is an indispensable part of the quantum
computation. In quantum computing especially the calculation model
of quantum circuit, the quantum logic gate is a basic operation of
qubits quantum circuit. Recently, a number of schemes have been
proposed to perform quantum logic gates and quantum information
processing using optical devices \cite{YFH2004}, quantum dot
\cite{BQH2002,CWX2015}, QED systems \cite{SBZ2013}, ion trap and
superconducting devices
\cite{JIC1995,WDJ2011,M2001,CPY2003,CPY2006}. As we know that a
universal set of quantum operations can be constructed by a series
of single- and two-qubit gates. Even though a multi-qubits gate
 could be decomposed into one- and two-qubit gates, it will be extremely
 complex and time-consuming in practical operation, that the arising decoherence
will destroy the quantum system eventually. An efficient method of
quantum information processing requires the whole operation is
robust against the decoherence, easily prepared and measured.
Adiabatic passage technique provides the robustness of the method
against small variations of field parameters, and the decoherence
caused by spontaneous emission can be avoided if the dynamics
follows dark states, i.e. states without components on lossy
excited states. However, that is not precise. If the controlling
parameters do not change slowly enough, the transitions between
different time-dependent instantaneous eigenstates may still
exist, that the fidelity of the evolved state with respect to the
target one must be reduced. In other words, the adiabatic passage
technique usually needs a long process \cite{JR1989}. If the
required evolution time is too long, the speed of the system
evolution will be slowed down, that the dissipation caused by
decoherence, noise, and losses would destroy the expected dynamics
finally.

Shortcuts to adiabatic passage is a promising technique for
quantum information processing which actually fight against the
decoherence, noise, or losses that are accumulated during a long
operation time. Thus, a variety of schemes have been proposed to
construct shortcuts to adiabatic passage in both theories and
experiment
\cite{ARXD2012,XASA2010,KPYR2011,AC2013,MYLJ2014,YYQJ2014,AFTS2012,JXPP2011,AC2012,HS2014}.
The shortcuts to adiabatic passage of logical gates operation have
been presented, too. Chen\emph{ et al} \cite{YHC2014} proposed a
scheme of shortcuts to adiabatic passage for performing a $\pi$
phase gate. And we have proposed a scheme of shortcut to adiabatic
passage for constructing the multiqubit controlled phase gate
\cite{YL2015}.

In this paper, we effectively combine the advantages of shortcuts
to adiabatic passage and quantum Zeno dynamic (QZD)
\cite{PHTA1995,PVGS2000} to implement a SWAP gate. It dose not
need the composition of element gates, but directly implements the
deterministic SWAP gate through designing resonant laser pulses by
the invariant-based inverse engineering. The logical SWAP gates in
our scheme can be performed in a much shorter time than that based
on adiabatic passage technique, and it is very robustness to the
decoherence caused by atomic spontaneous emission and cavity
decay.

~This paper is structured as follows: In section 2, we give a
brief description of the preliminary theory about Lewis-Riesenfeld
invariants and QZD. In section 3, we effectively combine the
shortcuts to adiabatic passage and QZD to implement the
deterministic SWAP gate. Section 4 shows the numerical simulation
results and feasibility analysis. The conclusion appears in
section 5.
\section{Preliminary theory}

\subsection{Lewis-Riesenfeld invariants}
We first make a brief introduction of the Lewis-Riesenfeld
invariants theory \cite{HRL1969,MAL2009}. Considering a system
which is governed by a time-dependent Hamiltonian $H(t)$, and we
can seek the time-dependent Hermitian invariants $I(t)$ that is
related to the original Hamiltonian $H(t)$ to satisfy
\begin{eqnarray}\label{1}
i\hbar \frac{\partial I(t)}{\partial t}&=&[H(t),I(t)].
\end{eqnarray}
Obviously, its expectation values remain constant all the time,
and drives the system state evolve along the initial eigenstate of
$I(t)$. For the time-dependent Schr\"odinger equation $i\hbar$
$\frac{\partial |\Psi(t)\rangle}{\partial t}$ $
=H(t)|\Psi(t)\rangle$, the solution can be expressed by the
superposition of ¡°dynamical modes¡± $|\Phi_{n} (t)\rangle$ of the
invariants $I(t)$
\begin{eqnarray}\label{2}
|\Psi(t)\rangle&=&\sum_n C_n e^{i\alpha_n}|\Phi_{n}(t)\rangle,
\end{eqnarray}
where $n=0,1,...$, and $C_n$ is one of the time-independent
amplitudes, $\alpha_n$ is the Lewis-Riesenfeld phase.
$|\Phi_{n}(t)\rangle$ is one of the orthonormal eigenvectors of
the invariant $I(t)$ with the corresponding real eigenvalue
$\lambda_n$, satisfying
$I(t)|\Phi_{n}(t)\rangle=\lambda_n|\Phi_{n}(t)\rangle$. And the
Lewis-Riesenfeld phase satisfies
\begin{eqnarray}\label{3}
\alpha_n(t)&=&\frac{1}{\hbar}\int_0^t dt^\prime
\langle\Phi_n(t^\prime )|i\hbar\frac{\partial}{\partial t^\prime
}-H(t^\prime )|\Phi_n(t^\prime )\rangle.
\end{eqnarray}

\subsection{Quantum Zeno dynamics}
The quantum Zeno effect was first proposed by Misra and Sudarshan,
and it can be used to reduce the influence of the decoherence and
dissipation by inhibiting the transition between quantum states
through the frequent measurements \cite{BM1977}. We consider a
system which is governed by the Hamiltonian
\begin{eqnarray}\label{4}
H_K=H_{\rm obs}+KH_{\rm meas},
\end{eqnarray}
 where
$H_{\rm obs}$ is the Hamiltonian of the investigated quantum
system and the $H_{\rm meas}$ is the interaction Hamiltonian
performing the measurement. $K$ is a coupling constant, and when
it satisfies $K\rightarrow \infty$, the whole system is governed
by the evolution operator
\begin{eqnarray}\label{5}
U(t)={\rm exp}[-it\sum_{n}(K\lambda_nP_n+P_nH_{\rm obs}P_n)],
\end{eqnarray}
where $P_n$ is one of the eigenprojections of $H_{\rm meas}$ with
eigenvalues $\lambda_n$($H_{\rm meas} = \sum_{n}\lambda_nP_n$).

\section{Shortcuts to adiabatic passage for the
deterministic SWAP gate}

\begin{figure}
\begin{center}
\includegraphics[scale=0.4]{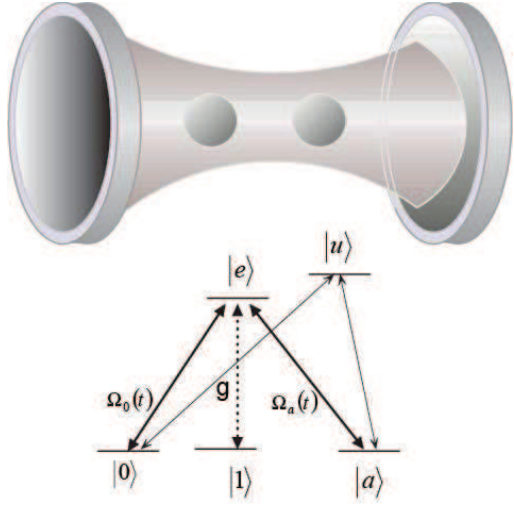}\caption{(Color online) The schematic setup of SWAP gate implementation. The
atoms are trapped in a single mode optical cavity, and each atom
possesses five atomic levels. }
\end{center}
\end{figure}

We consider a system in which the atoms are fixed inside an
optical cavity as shown in figure 1. Each atom possesses three
ground states $|0\rangle$, $|1\rangle$, $|a\rangle$ and two
excited states $|e\rangle$, $|u\rangle$. The transitions
$|0\rangle\leftrightarrow|e\rangle$ and
$|a\rangle\leftrightarrow|e\rangle$ are coupled to the laser
pulses, with the corresponding Rabi frequencies $\Omega_{0}(t)$
and $\Omega_{a}(t)$, respectively. The transition
 $|1\rangle\leftrightarrow|e\rangle$ is strongly coupled to the single mode cavity
field with the coupling constant $g$. And the auxiliary excited
state $|u\rangle$ is only used to implement the one-qubit
operation.

Now we show the scheme to realize a deterministic SWAP gate. The
initial state $|\Psi_{0}\rangle$ of the two atoms and the cavity
field is defined as
\begin{eqnarray}\label{6}
|\Psi_0\rangle&=&\alpha_{00}|00\rangle_{AB}|0\rangle_{C}+\alpha_{01}|01\rangle_{AB}|0\rangle_{C}\cr
&&
+\alpha_{10}|10\rangle_{AB}|0\rangle_{C}+\alpha_{11}|11\rangle_{AB}|0\rangle_{C},
\end{eqnarray}
where $|nm\rangle_{AB}|0\rangle_{C}$ denotes atoms $A$ and $B$ are
in the state $|nm\rangle_{AB}$ ($n,m=0,1$), and the cavity is in
vacuum state $|0\rangle_{C}$. $\alpha_{nm}$ denotes the amplitude
of the state $|nm\rangle_{AB}|0\rangle_{C}$, and satisfies the
normalization condition. The SWAP gate makes the values of the two
qubits exchange and then the output state becomes
\begin{eqnarray}\label{7}
|\Psi\rangle&=&\alpha_{00}|00\rangle_{AB}|0\rangle_{C}+\alpha_{01}|10\rangle_{AB}|0\rangle_{C}\cr
&&
+\alpha_{10}|01\rangle_{AB}|0\rangle_{C}+\alpha_{11}|11\rangle_{AB}|0\rangle_{C}.
\end{eqnarray}

There are three steps for the implementation of SWAP gate. First,
we transfer the population of $|01\rangle_{AB}|0\rangle_{C}$ to
$-|1a\rangle_{AB}|0\rangle_{C}$ completely with the help of laser
pulses resonant with $A$ atomic transition
$|0\rangle_{A}\leftrightarrow|e\rangle_{A}$ and $B$ atomic
transition $|a\rangle_{B}\leftrightarrow|e\rangle_{B}$ with the
corresponding Rabi frequencies $\Omega_{0A}(t)$ and
$\Omega_{aB}(t)$, respectively.  After the interaction, the
initial state becomes
\begin{eqnarray}\label{8}
|\Psi_1\rangle&=&\alpha_{00}|00\rangle_{AB}|0\rangle_{C}-\alpha_{01}|1a\rangle_{AB}|0\rangle_{C}\cr
&&
+\alpha_{10}|10\rangle_{AB}|0\rangle_{C}+\alpha_{11}|11\rangle_{AB}|0\rangle_{C}.
\end{eqnarray}

Then the population of $|10\rangle_{AB}|0\rangle_{C}$ is
completely transferred to $-|a1\rangle_{AB}|0\rangle_{C}$ with the
similar method with the help of laser pulses resonant with $B$
atomic transition $|0\rangle_{B}\leftrightarrow|e\rangle_{B}$ and
$A$ atomic transition $|a\rangle_{A}\leftrightarrow|e\rangle_{A}$
with the corresponding Rabi frequencies $\Omega_{0B}(t)$ and
$\Omega_{aA}(t)$, respectively. And the state of the system
becomes
\begin{eqnarray}\label{9}
|\Psi_2\rangle&=&\alpha_{00}|00\rangle_{AB}|0\rangle_{C}-\alpha_{01}|1a\rangle_{AB}|0\rangle_{C}\cr
&&
-\alpha_{10}|a1\rangle_{AB}|0\rangle_{C}+\alpha_{11}|11\rangle_{AB}|0\rangle_{C}.
\end{eqnarray}

Finally the population of the single atom state $|a\rangle_{A(B)}$
is transferred into $-|0\rangle_{A(B)}$ by the one-qubit operation
with the help of the auxiliary excited state $|u\rangle_{A(B)}$,
and the corresponding laser pulses are resonant with $A(B)$ atomic
transition $|0\rangle_{A(B)}\leftrightarrow|u\rangle_{A(B)}$ and
$|a\rangle_{A(B)}\leftrightarrow|u\rangle_{A(B)}$ with the
corresponding Rabi frequencies $\Omega_{aA(B)}^{'}(t)$ and
$\Omega_{0A(B)}^{'}(t)$, respectively. As a result, the state of
the system turns into
\begin{eqnarray}\label{10}
|\Psi_3\rangle&=&\alpha_{00}|00\rangle_{AB}|0\rangle_{C}+\alpha_{01}|10\rangle_{AB}|0\rangle_{C}\cr
&&
+\alpha_{10}|01\rangle_{a1}|0\rangle_{C}+\alpha_{11}|11\rangle_{AB}|0\rangle_{C},
\end{eqnarray}
thus, the SWAP gate is obtained.

In the following, we show the details of how to realize the SWAP
gate by combining the shortcuts to adiabatic passage and QZD. For
the first step, the input laser pulses are resonant with $A$
atomic transition $|0\rangle_{A}\leftrightarrow|e\rangle_{A}$ and
 $B$ atomic transition
$|a\rangle_{B}\leftrightarrow|e\rangle_{B}$ with the corresponding
Rabi frequencies $\Omega_{0A}(t)$ and $\Omega_{aB}(t)$,
respectively. The Hamiltonian of the first step in the interaction
picture can be written as ($\hbar=1$)
\begin{eqnarray}\label{11}
H_{\rm I}&=&H_{l}+H_{c},
\end{eqnarray}
\begin{eqnarray}\label{12}
H_{l}&=&\Omega_{0A}(t)|e\rangle_{A}\langle{0}|+\Omega_{aB}(t)|e\rangle_{B}\langle{a}|
 +\rm H.c.,
\end{eqnarray}
\begin{eqnarray}\label{13}
H_{c}&=&g_{A}a|e\rangle_{A}\langle{1}|+g_{B}a|e\rangle_{B}\langle{1}|
+\rm H.c.,
\end{eqnarray}
here $a$ is the annihilation operator, and $g_{A(B)}$ is the
coupling strength between cavity mode and the trapped atom. In the
following, we will set $g_{A}=g_{B}$ for simplicity. Since the
lasers do not couple with the atomic state $|1\rangle$, that the
states $|11\rangle_{AB}|0\rangle_{C}$ and
$|10\rangle_{AB}|0\rangle_{C}$ are decoupled from the other states
in this step, and will not participate the evolution any more. For
the state $|00\rangle_{AB}|0\rangle_{C}$, the evolution subspace
can be spanned by the basis vectors
$|\varphi_1\rangle=|00\rangle_{AB}|0\rangle_C,$
$|\varphi_2\rangle=|e0\rangle_{AB}|0\rangle_C,$ and
$|\varphi_3\rangle=|10\rangle_{AB}|1\rangle_C.$ The Hamiltonian of
this system reads
\begin{equation}
H_{1}=\Omega_{0A}(t)|e\rangle_{A}\langle0|+g_{A}|e\rangle_{A}\langle1|+\rm{H.c}
.
\end{equation}
For the condition $g_{A}\gg\Omega_{0A}(t)$, the Hilbert space is
 split into three invariant subspaces
 $\Gamma_{1}= |\varphi_1\rangle $,
 $\Gamma_{2}=\frac{1}{\sqrt{2}}(-|\varphi_2\rangle+|\varphi_3\rangle)$
 and
 $\Gamma_{3}=\frac{1}{\sqrt{2}}(|\varphi_2\rangle+|\varphi_3\rangle)$
 with the eigenvalues $\lambda_{1}=0$, $\lambda_{2}=-g_{A}$, and
 $\lambda_{3}=g_{A}$. The
 corresponding projections are $\rm
 P_{n}^{\alpha}=|\alpha\rangle\langle\alpha|$
 ($|\alpha\rangle\in\Gamma_{n}, n=1,2,3$). According to the QZD, $U(t)={\rm exp}[-it\sum_{n}(K\lambda_nP_n+P_nH_{\rm
 obs}P_n)]$, the effective Hamiltonian reduces to
\begin{equation}
H_{\rm eff}=g_{A}|e\rangle_{A}\langle1|+g_{A}|1\rangle_{A}\langle
e|,
\end{equation}
which indicates that the transition between
$|00\rangle_{AB}|0\rangle_{C}$ and $|e0\rangle_{AB}|0\rangle_{C}$
is eliminated. For another state $|01\rangle_{AB}|0\rangle_{C}$,
the system evolves in the subspace which is
 spanned by the basis vectors $|\phi_1\rangle=|01\rangle_{AB}|0\rangle_C,$
$|\phi_2\rangle=|e1\rangle_{AB}|0\rangle_C,$
$|\phi_3\rangle=|11\rangle_{AB}|1\rangle_C,$
$|\phi_4\rangle=|1e\rangle_{AB}|0\rangle_C,$ and
$|\phi_5\rangle=|1a\rangle_{AB}|0\rangle_C.$ This evolution system
is governed by the Hamiltonian
\begin{eqnarray}\label{16}
H_{2}&=&\Omega_{0A}(t)|e\rangle_{A}\langle0|+g_{A}|e\rangle_{A}\langle1|\cr
&&+\Omega_{aB}(t)|e\rangle_{B}\langle
a|+g_{B}|e\rangle_{B}\langle1|+\rm{H.c}.
\end{eqnarray}
By using the similar way to the above processes, the effective
Hamiltonian of this system reads
\begin{eqnarray}\label{17}
H_{\rm
eff}^{'}=\frac{1}{\sqrt{2}}|\mu\rangle(\Omega_{0A}(t)\langle\phi_1|+\Omega_{aB}(t)\langle\phi_5|+
\rm{H.c}),
\end{eqnarray}
where
$|\mu\rangle=\frac{1}{\sqrt{2}}(-|\phi_2\rangle+|\phi_4\rangle)$.
To speed up the transition from $|\phi_1\rangle$ to
$-|\phi_5\rangle$ by the invariant-based inverse engineering, we
must find out the invariant Hermitian operator $I(t)$ which
satisfies (\ref{1}). Since the effective Hamiltonian $H_{\rm
eff}^{'}$ possesses the SU(2) dynamical symmetry, the invariant
$I(t)$ can be given by \cite{XCE2011}
\begin{eqnarray}\label{18}
I(t)&=&\frac{1}{\sqrt{2}}\chi(\cos\nu\sin\beta|\mu\rangle\langle\phi_1|+\cos\nu\cos\beta|\mu\rangle\langle\phi_5|\cr
&&+i\sin\nu|\phi_5\rangle\langle\phi_1|+\rm{H.c} ),
\end{eqnarray}
where $\chi$ is an arbitrary constant with units of frequency to
keep $I(t)$ with dimensions of energy. The time-dependent
auxiliary parameters $\nu$ and $\beta$ satisfy the equations
\begin{eqnarray}\label{19}
\dot{\nu}&=&\frac{1}{\sqrt{2}}(\Omega_{0A}(t)\cos\beta-\Omega_{aB}(t)\sin\beta),\nonumber\\
\dot{\beta}&=&\frac{1}{\sqrt{2}}\tan\nu(\Omega_{aB}(t)\cos\beta+\Omega_{0A}(t)\sin\beta).
\end{eqnarray}
From (\ref{19}) the expressions of $\Omega_{0A}(t)$ and
$\Omega_{0B}(t)$ can be derived as follow
\begin{eqnarray}\label{20}
\Omega_{0A}(t)&=&\sqrt{2}(\dot{\beta}\cot\nu\sin\beta+\dot{\nu}\cos\beta),\nonumber\\
\Omega_{aB}(t)&=&\sqrt{2}(\dot{\beta}\cot\nu\cos\beta-\dot{\nu}\sin\beta),
\end{eqnarray}
where the dot denotes a time derivative. The eigenstates of the
invariant $I(t)$ are
\begin{eqnarray}\label{21}
|\Phi_0(t)\rangle&=&\cos\nu\cos\beta|\phi_1\rangle-i\sin\nu|\mu\rangle-\cos\nu\sin\beta|\phi_5\rangle,\nonumber\\
|\Phi_\pm(t)\rangle&=&\frac{1}{\sqrt{2}}[(\sin\nu\cos\beta\pm
i\sin\beta)|\phi_1\rangle+i\cos\nu|\mu\rangle\cr
&&-(\sin\nu\sin\beta\mp i\cos\beta)|\phi_5\rangle],
\end{eqnarray}
with the eugenvalues $\varepsilon_0=0$ and $\varepsilon_\pm=\pm1,$
respectively. The solution of the Schr\"{o}dinger equation
$i\hbar$ $\frac{\partial |\Psi(t)\rangle}{\partial t}$ $
=H(t)|\Psi(t)\rangle$ can be written as the superposition of the
eigenstates of $I(t)$
\begin{eqnarray}\label{22}
|\Psi(t)\rangle&=&\sum_{n=0,\pm}C_ne^{i\alpha_n}|\Phi_n(t)\rangle,
\end{eqnarray}
where $\alpha_n(t)$ is the  Lewis-Riesenfeld phase in (3), and
$C_n$ is a time-independent amplitude. In order to get the final
state $-|\phi_5\rangle$, we choose the parameters appropriately
\begin{eqnarray}\label{23}
\nu(t)&=&\epsilon,~~~~~~~~~\beta(t)=\frac{\pi t}{2t_f},
\end{eqnarray}
where $\epsilon$ is a time-independent small value. From the
precise calculation, we can easily obtain
\begin{eqnarray}\label{24}
\Omega_{0A}(t)&=&\frac{\pi}{\sqrt{2}t_f}\cot\epsilon\sin\frac{\pi t}{2t_f},\nonumber\\
\Omega_{aB}(t)&=&\frac{\pi}{\sqrt{2}t_f}\cot\epsilon\cos\frac{\pi
t}{2t_f}.
\end{eqnarray}
When $t=t_f$,
\begin{eqnarray}\label{25}
|\Psi(t_f)\rangle&=&-\sin\epsilon\sin\alpha|\phi_1\rangle\cr
&&+(-i\sin\epsilon\cos\epsilon+i\sin\epsilon\cos\epsilon\cos\alpha)|\mu\rangle\cr
&&+(-\cos^2\epsilon-\sin^2\epsilon\cos\alpha)|\phi_5\rangle,
\end{eqnarray}
where $\alpha=\pi/(2\sin\epsilon)=|\alpha_\pm|$, and $\alpha_\pm$
are the Lewis-Riesenfeld phases. As long as $\alpha$ satisfies the
condition $\alpha=2N\pi(N=1,2,3...)$,
$|\Psi(t_f)\rangle=-|\phi_5\rangle=-|1a\rangle_{AB}|0\rangle_C$
can be achieved. Then the first step is realized successfully, and
$|\Psi_1\rangle =
\alpha_{00}|00\rangle_{AB}|0\rangle_{C}-\alpha_{01}|1a\rangle_{AB}|0\rangle_{C}
+\alpha_{10}|10\rangle_{AB}|0\rangle_{C}+\alpha_{11}|11\rangle_{AB}|0\rangle_{C}$
can be obtained.

The second step is just similar to the first step. In this step,
the laser pulses are resonant with $B$ atomic transition
$|0\rangle_{B}\leftrightarrow|e\rangle_{B}$ and $A$ atomic
transition $|a\rangle_{A}\leftrightarrow|e\rangle_{A}$ with the
corresponding Rabi frequencies
$\Omega_{0B}(t)=\frac{\pi}{\sqrt{2}t_f}\cot\epsilon\sin\frac{\pi
t}{2t_f}$ and
$\Omega_{aA}(t)=\frac{\pi}{\sqrt{2}t_f}\cot\epsilon\cos\frac{\pi
t}{2t_f}$, respectively. In this step, the states
$|00\rangle_{AB}|0\rangle_C$, $-|1a\rangle_{AB}|0\rangle_C$ and
$|11\rangle_{AB}|0\rangle_C$ will be remained the same, while the
population of the state $|10\rangle_{AB}|0\rangle_C$ is completely
transferred to the state $-|a1\rangle_{AB}|0\rangle_C$, then
$|\Psi_2\rangle =
\alpha_{00}|00\rangle_{AB}|0\rangle_{C}-\alpha_{01}|1a\rangle_{AB}|0\rangle_{C}
-\alpha_{10}|a1\rangle_{AB}|0\rangle_{C}+\alpha_{11}|11\rangle_{AB}|0\rangle_{C}$
can be achieved.

The third step is just a one-qubit operation with the help of the
auxiliary state $|u\rangle_{A(B)}$. The input laser pulses are
resonant with $A(B)$ atomic transition
$|0\rangle_{A(B)}\leftrightarrow|u\rangle_{A(B)}$ and
$|a\rangle_{A(B)}\leftrightarrow|u\rangle_{A(B)}$ with the
corresponding Rabi frequencies
$\Omega_{aA(B)}(t)^{'}=\frac{\pi}{\sqrt{2}t_f}\cot\epsilon\sin\frac{\pi
t}{2t_f}$ and
$\Omega_{0A(B)}^{'}(t)=\frac{\pi}{\sqrt{2}t_f}\cot\epsilon\cos\frac{\pi
t}{2t_f}$, respectively. The Hamiltonian in this step is given as
follow
\begin{eqnarray}\label{26}
H_{3}=\sum_{i=A,B}\Omega_{ai}^{'}(t)|u\rangle_{i}\langle
a|+\Omega_{0i}^{'}(t)|u\rangle_{i}\langle0|+ \rm{H.c}.
\end{eqnarray}
For the initial atomic state $|a\rangle_{A(B)}$, the population
will finally transfer to the state $-|0\rangle_{A(B)}$ by using
the similar way in the first step. Thus, we can obtain
$|\Psi_3\rangle=\alpha_{00}|00\rangle_{AB}|0\rangle_{C}+\alpha_{01}|10\rangle_{AB}|0\rangle_{C}
+\alpha_{10}|01\rangle_{a1}|0\rangle_{C}+\alpha_{11}|11\rangle_{AB}|0\rangle_{C}$.
That is the result of the SWAP gate. figure 2 represents these
three steps for constructing the SWAP gate.

\begin{figure}\centering
\includegraphics[width=3.3in,height=4.5in]{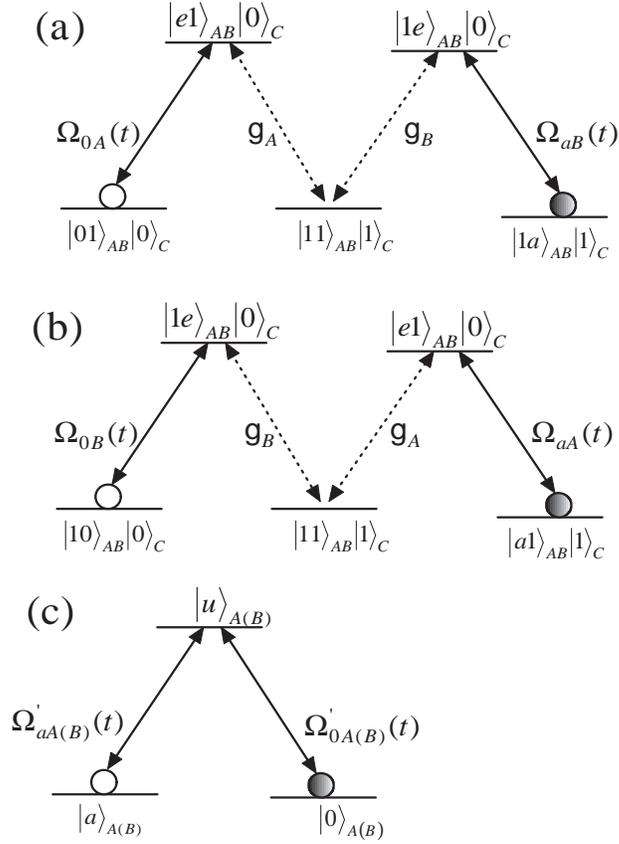}\caption{(Color online) The three steps for constructing the SWAP gate. For each
step, the initial state is denoted by an empty circle and the
final state is represented by a black circle. }
\end{figure}

\begin{figure} \centering
\includegraphics[width=3.5in,height=1.5in]{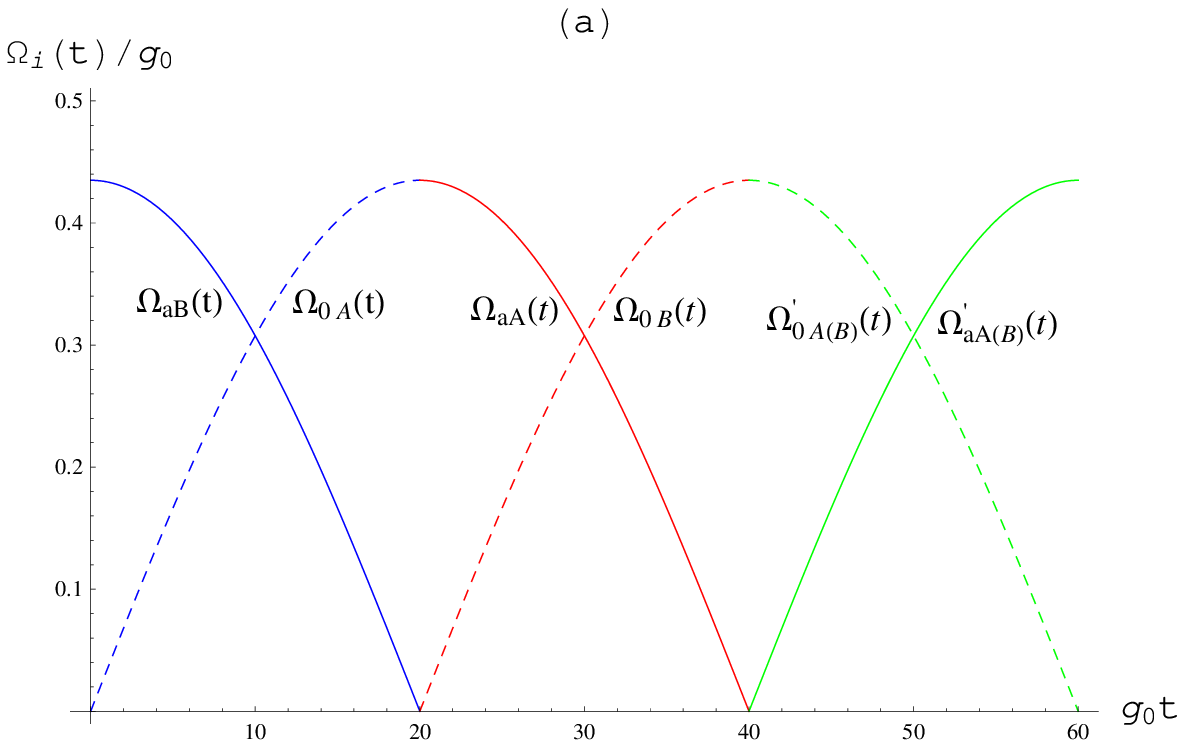}%
\hspace{0.3in}%
\includegraphics[width=3.5in,height=1.5in]{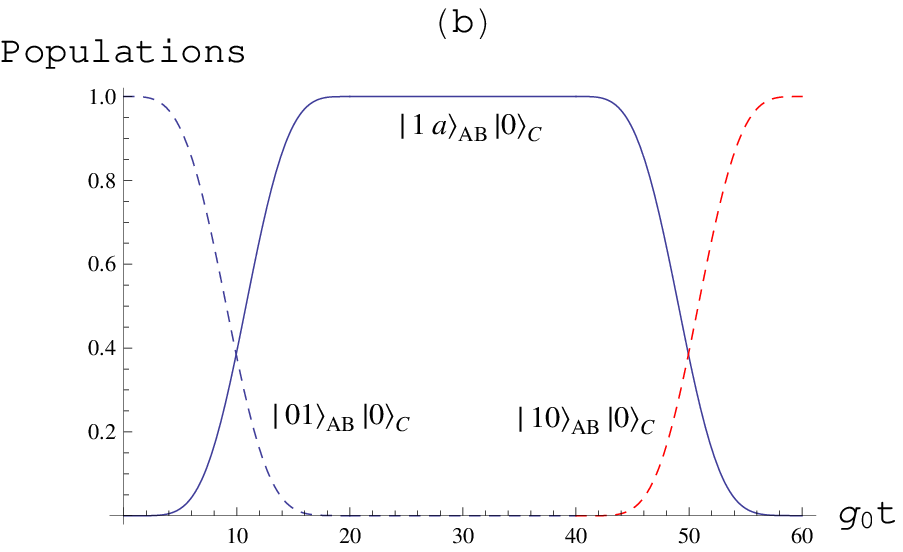}%
\hspace{0.3in}%
\includegraphics[width=3.5in,height=1.5in]{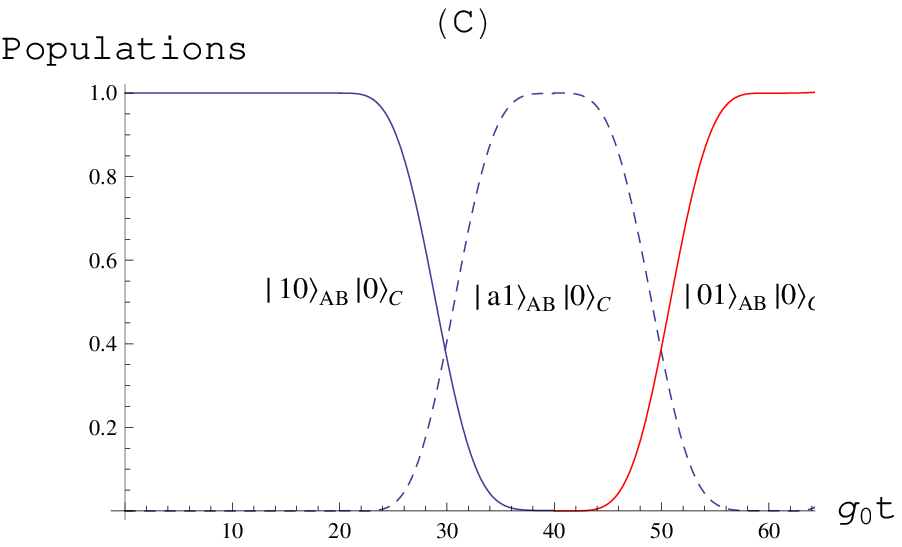}\caption{(Color online) (a) Temporal profile of the time dependence Rabi
frequencies $\Omega_{i}(t)/g_{0}$ versus $t/g_{0}$ with
$\Omega_i(t)=\Omega_{0A}(t)$ (dash blue line), $\Omega_{aB}(t)$
(solid blue line), $\Omega_{0B}(t)$ (dash red line),
$\Omega_{aA}(t)$ (solid red line), $\Omega_{aA(B)}^{'}(t)$ (solid
green line), $\Omega_{0A(B)}(t)^{'}$ (dash green line). (b) Time
evolutions of the populations of the corresponding system states
with the initial states $|01\rangle$. (c) Time evolutions of the
populations of the corresponding system states with the initial
state $|10\rangle$. The system parameters are set to be $\epsilon=
0.25$, $g_{A}=g_{B}=10g_{0}$ and $ t_{f}=20/g_{0}.$}
\end{figure}

\begin{figure}\centering
\includegraphics[width=3.5in]{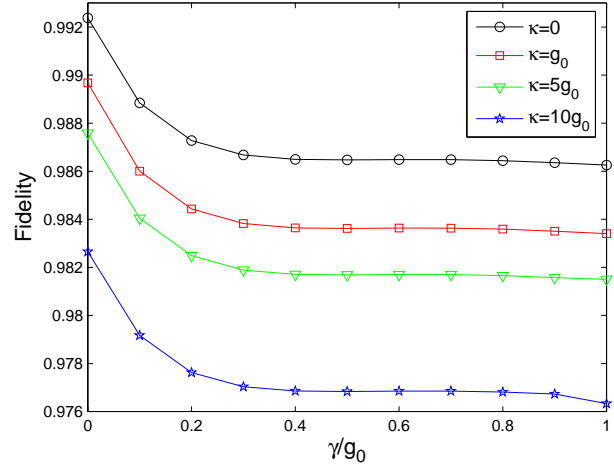}\caption{(Color online) The effect of atomic spontaneous emission $\gamma$ on the
fidelity of the SWAP gate with different values of the cavity
decay $\kappa$.}
\end{figure}
\section{Numerical simulations and feasibility analysis}
In the following, we make the numerical simulations to verify the
validity of the SWAP gate. figure 3(a) shows the time-dependence
laser pulse $\Omega_{m}(t)/g_{0}$ as a function of $t/g_{0}$ for a
fixed value $\epsilon= 0.25$, $g_{A}=g_{B}=10g_{0}$ and
$t_{f}=20/g_{0}$. With these parameters, the Zeno condition can be
met well. The populations of the states
$|01\rangle_{AB}|0\rangle_{C}$ and $|10\rangle_{AB}|0\rangle_{C}$
swap perfectly which can be seen from figure 3(a) and figure 3(b).
Whether a scheme is available largely depends on the robustness to
the loss and decoherence, so in the following, we consider the
loss and decoherence effects on our SWAP gate operation. The
corresponding master equation for the whole system density matrix
$\rho(t)$ has the following form:
\begin{eqnarray}\label{27}
\dot{\rho(t)}&=&-i[H_{\rm
{total}},\rho(t)]-\frac{\kappa}{2}[a^{\dag}a\rho(t)-2a\rho(t)a^{\dag}+\rho(t)a^\dag
a ]\cr
&&-\frac{\gamma_A}{2}\sum_{l=e,u}\sum_{k=0,1,a}[\sigma_{l,l}^A\rho(t)-2\sigma_{k,l}^A\rho(t)\sigma_{l,k}^A+\rho(t)\sigma_{l,l}^A]\cr
&&-\frac{\gamma_B}{2}\sum_{l=e,u}\sum_{k=0,1,a}[\sigma_{l,l}^B\rho(t)-2\sigma_{k,l}^B\rho(t)\sigma_{l,k}^B+\rho(t)\sigma_{l,l}^B],\nonumber\\
\end{eqnarray}
where $H_{\rm {total}}=H_{1}+H_{2}+H_{3}$. $\kappa$ is the cavity
decay rate, $\gamma_{A(B)}$ is $A(B)$ atomic spontaneous emission
rate from the excited state $|l\rangle_{i}(l=e,u;i=A,B)$ to the
ground state $|k\rangle_{i}(k=0,1,a)$, respectively.
$\sigma_{l,k}=|l\rangle\left\langle k\right|$. For simplicity, we
assume $\gamma_A=\gamma_B=\gamma$ and the initial condition
$\rho(0)= |\Psi_0\rangle\left\langle \Psi_0\right|$. In figure 4,
the fidelity of the SWAP gate is plotted versus the dimensionless
parameter $\gamma/g$ with different values of $\kappa/g_{0}$ by
numerically solving the master (\ref{27}). From figure 4 we can
see that the fidelity of our SWAP gate is higher than $97.6\%$
even when the values of $\gamma$ and $\kappa$ are comparable to
$\gamma=0.1g_{A(B)}=g_{0}$ and $\kappa=g_{A(B)}=10g_{0}$. It shows
that the SWAP gate in our scheme is robust against decoherence due
to cavity decay and atomic spontaneous emission.

Now we give a brief analysis of the feasibility in experiment for
this scheme. The scheme can be realized with trapped ions and
nitrogen-vacancy color center in diamond \cite{MSS2002}, cavity
QED systems \cite{SSD2000,JPH2002,LYX2003} or with impurity levels
in a solid, such as Pr$^{3+}$ ions in Y$_2$SiO$_5$ crystal
\cite{KI2001}. In recent experiments, the fabrication of various
high-Q microcavities including whispering-gallery-mode cavities
\cite{DK2003,SM2005}, micropost cavities, and one- or
two-dimensional photonic-crystal microcavities
\cite{BS2005,TT2007} also have been well designed. And the
suitable parameters of toroidal microcavity system for
strong-coupling cavity QED have been investigated and the
parameters can be chosen as
$(g,\kappa,\gamma)/2\pi=(750,3.5,2.62)$ MHz \cite{SM2005}. In this
case, $\gamma \simeq 0.0034g_{A(B)}=0.034g_{0}$, $\kappa
\simeq0.0046g_{A(B)}=0.046g_{0}$, and with these parameters the
fidelity of SWAP gate can reach $99\%$. So our scheme is robust
against both the cavity decay and atomic spontaneous emission and
could be very promising and useful for quantum information
processing.

\section{Conclusion}
In summary, we have proposed a promising scheme to implement a
SWAP gate through the shortcut to adiabatic passage and QZD
instead of relying on the compositions of a large number of
elementary gates. We also study the influences of atomic
spontaneous emission and cavity decay on the fidelity through
numerical simulation. The numerical simulation results denote that
our scheme is very robust against the decoherence caused by atomic
spontaneous emission and cavity decay, so it can be a more
reliable choice in experiment. We believe that our scheme will be
useful to realize quantum algorithms, such as Shor's algorithm for
prime factoring or Grover's algorithm for database search.

\ack

 This work was supported by the National Natural Science
Foundation of China under Grant Nos. 11464046, 11264042, 11465020
and 61465013.

\end{document}